\documentclass{PoS}

\usepackage{amsmath}
\usepackage{amssymb}
\usepackage{slashed}
\usepackage{listings}
\usepackage{cite}

\newcommand{\galph}{\ensuremath{G_\alpha^{(1)}}}
\newcommand{\gao}[1]{\ensuremath{G_{\mathrm{A}_k}^{(#1)}}}
\newcommand{\gvo}[1]{\ensuremath{G_{\mathrm{V}_0}^{(#1)}}}
\newcommand{\halph}{\ensuremath{H_\alpha^{(1)}}}
\newcommand{\hvo}[1]{\ensuremath{H_{\mathrm{V}_0}^{(#1)}}}
\newcommand{\hao}[1]{\ensuremath{H_{\mathrm{A}_k}^{(#1)}}}

\renewcommand{\vec}[1]{\mathbf{{#1}}}

\newcommand{\ord}[1]{^{(#1)}}

\newcommand{\of}[1]{{\left[#1\right]}}

\newcommand{\avg}[1]{\langle #1 \rangle}

\newcommand{\pastor}{{\texttt{pastor}}}
\newcommand{\msbar}{{\overline{M\!S}}}
\newcommand{\lat}{{\mathrm{lat}}}

\newcommand{\zmlat}{\ensuremath{Z_{\mathrm{m},\lat}}}

\newcommand{\za}{\ensuremath{Z_{\mathrm{A}}}}

\newcommand{\zv}{\ensuremath{Z_{\mathrm{V}}}}
\newcommand{\zva}{\ensuremath{Z_{\mathrm{V/A}}}}

\newcommand{\bv}{\ensuremath{b_{\mathrm{V}}}}
\newcommand{\ba}{\ensuremath{b_{\mathrm{A}}}}

\newcommand{\CP}{\ensuremath{C_{\mathrm{PS}}}}
\newcommand{\CV}{\ensuremath{C_{\mathrm{V}}}}
\newcommand{\rgi}{\ensuremath{{\mathrm{RGI}}}}
\newcommand{\mr}{\ensuremath{\overline{m}}}

\newcommand{\Bv}{\ensuremath{B_{\mathrm{V}}^\stat}}
\newcommand{\Ba}{\ensuremath{B_{\mathrm{A}}^\stat}}

\newcommand{\fov}{\ensuremath{f_1^{\mathrm{V}_0}}}
\newcommand{\foa}{\ensuremath{f_1^{\mathrm{A}_1}}}

\newcommand{\phiv}{\ensuremath{\Phi^{\mathrm{V}_0}}}
\newcommand{\phia}{\ensuremath{\Phi^{\mathrm{A}_k}}}
\newcommand{\phivs}{\ensuremath{\Phi^{\mathrm{V}_0,\stat}}}
\newcommand{\phias}{\ensuremath{\Phi^{\mathrm{A}_k,\stat}}}

\newcommand{\xabare}{\ensuremath{X^\mathrm{bare}_{\mathrm{A}_k}}}
\newcommand{\xvbare}{\ensuremath{X^\mathrm{bare}_{\mathrm{V}_0}}}
\newcommand{\xa}{\ensuremath{X_{\mathrm{A}_k}}}
\newcommand{\xv}{\ensuremath{X_{\mathrm{V}_0}}}
\newcommand{\xao}[1]{\ensuremath{X_{\mathrm{A}_k}^{(#1)}}}
\newcommand{\xvo}[1]{\ensuremath{X_{\mathrm{V}_0}^{(#1)}}}
\newcommand{\xabareo}[1]{\ensuremath{X^{\mathrm{bare},(#1)}_{\mathrm{A}_k}}}

\newcommand{\xvbareo}[1]{\ensuremath{X^{\mathrm{bare},(#1)}_{\mathrm{V_0}}}}

\newcommand{\psibar}{\ensuremath{\overline \psi{}}}

\newcommand{\zetabar}{\ensuremath{\overline \zeta{}}}

\newcommand{\mq}{m_\mathrm{q}}

\newcommand{\mc}{m_\mathrm{c}}

\newcommand{\bm}{b_\mathrm{m}}

\newcommand{\stat}{\mathrm{stat}}

\newcommand{\hqet}{\mathrm{HQET}}
\newcommand{\qcd}{\mathrm{QCD}}

\newcommand{\mh}{m_\mathrm{h}}

\title{HQET Flavor Currents Using Automated Lattice Perturbation Theory}

\ShortTitle{HQET Flavor Currents}

\author{\speaker{Dirk Hesse}\\
  Universit\'a degli studi di Parma, Viale G.P.\ Usberti 7/a, 43100
  Parma, Italy\\
  NIC, DESY, Platanenallee 6, 15738 Zeuthen, Germany\\
  E-mail: \email{dirk.hesse@fis.unipr.it}}

\abstract{ Matrix elements of heavy-light flavor currents play an
  important role in modern particle physics and precise theory
  predictions are of interest for phenomenology. Heavy Quark Effective
  Theory (HQET) is a valuable tool to obtain such predictions. In the
  HQET matching program of the ALPHA collaboration presently only the
  temporal component of the axial vector current is
  included. Extending the matching to the temporal component of the
  vector current and the spatial components of the axial vector
  current thus seems desirable. Here we present a recent one-loop
  study in lattice perturbation theory to test two candidate matching
  observables for these currents for their quality to guide future
  non-perturbative investigations \cite{paper}.}

\FullConference{The 30 International Symposium on Lattice Field Theory
  - Lattice 2012,\\
  June 24-29, 2012\\
  Cairns, Australia}

\begin{document}

\section{Introduction}

Matrix elements of heavy-light flavor currents are important
theoretical input to constrain physics of and beyond the standard
model through decays of heavy mesons. A theoretically clean way to
treat systems with one single heavy quark is Heavy Quark Effective
Theory (see \cite{Sommer:2010ic} and references therein), short HQET.
From the theory point of view, especially leptonic decays of heavy
mesons are relatively easy to treat. Hence next to the parameters of
the action, the HQET program of the ALPHA collaboration
\cite{Heitger:2003nj, Blossier:2010jk, Blossier:2010vz,
  Blossier:2010mk, DellaMorte:2006sv, Blossier:2012qu} included the
temporal component of the heavy-light axial vector current form the
very beginning.

Because of the current big interest in flavor physics within the
community, it is a logical next step to include all components of the
vector and axial vector current to make theory predictions for matrix
elements such as for example the one entering in the exclusive
semi-leptonic decay $B\to \pi l \nu$.

\subsection{HQET and matching}

HQET is a systematic expansion of QCD with one heavy quark in inverse
powers of its mass. In addition to the parameters in the HQET action,
one has to take care of the coefficients of the currents under
investigation as well. The inclusion of all components of the vector
and axial vector current increases the numbers of HQET parameters to
be fixed at order $1/m$ to a total of 19.  Considering this rather big
number of matching conditions, it seems useful to test them first in
relatively cheap perturbative calculations. This possibility is even
more attractive since we have the \pastor\ software package
\cite{Hesse:2011zz} for automated lattice perturbation theory
calculations at our disposal. Even though perturbative one-loop
calculations performed with \pastor\ only require a moderate effort we
do not intend to perform the matching of all 19 parameter
here. Instead we will focus on the renormalization constants of the
temporal component of the vector and the spatial components of the
axial vector current. These play an important role since as stated
above they were not included in the matching by the ALPHA
collaboration so far and enter already at the lowest order in HQET.

The matching procedure consists basically of evaluating the same
number of observables $\Phi_i$ as the number of parameters to be fixed
-- in QCD and HQET -- and setting $\Phi_i^\hqet = \Phi_i^\qcd$ to
determine (in fact define) the HQET parameters. The main objective of
this work will be to investigate possible matching conditions for the
currents mentioned above and estimate the size of higher order
contributions in the $1/m$-expansion. For a precise determination of
the parameters, one certainly wants these higher order effects to play
as little a role as possible.


\section{Matching of flavor currents}
We work in the Schr\"odinger functional
\cite{Luscher:1992an,Sint:1993un} with two mass-less quark flavors,
$m_1 = m_2 = 0$, and a third heavy one whose renormalized mass is as
usual given by the dimension-less parameter $z = \mr_3 L = \mr L$. The
boundary fields are given by $\zeta_i,\zetabar_i$ and the phase angle
for all three flavors is equal and denoted $\theta$. We are interested
in the temporal component of the vector current and the spatial
components of the axial vector current of a heavy and a light quark,
\begin{equation}
  \label{eq:c50}
  A_k(x) =\psibar_3(x) \gamma_k \gamma_5 \psi_2(x),\quad
  V_0 (x) = \psibar_3(x) \gamma_0 \psi_2(x),
\end{equation}
or to be more precise their matrix elements between heavy-light
pseudoscalar states. To this end, we define in a first step the
three-point functions (with implicit improvements $\ba,\bv$
\cite{Luscher:1996sc})
\begin{align}
 \foa (x_0; \theta, z) &\;= - 
 \frac {a^3} 2
  \sum_{\vec x} \avg{\zetabar_1\,\gamma_5\,\zeta_3 \, A_0(x)\, 
    \zetabar_2'\,\gamma_1 \,\zeta_1'},\quad \zeta_i = \frac{a^3}{L^{-3/2}} \sum_{\vec x} \zeta_i(\vec x),\quad 
  \zetabar_i = \frac{a^3} {L^{-3/2}} \sum_{\vec x} \zetabar_i(\vec
  x),\nonumber\\
  \fov (x_0; \theta, z) &\;= - 
 \frac {a^3} 2
  \sum_{\vec x} \avg{\zetabar_1\,\gamma_5\,\zeta_3 \, A_0(x)\, 
    \zetabar_2'\,\gamma_5 \,\zeta_1'}.   \label{eq:178}
\end{align}
To define observables that are suitable to extract the desired matrix
elements \cite{Heitger:2004gb}, we form the ratios
\begin{gather}
  \label{eq:247}
  \phiv(L, \mr) = \log \left(-
  \frac {\zv \,\fov(T/2; \theta,  z)}{\left[ f_1(\theta, z)
      f_1(\theta,0) \right ]^{1/2}}\right) ,\quad
   \phia(L, \mr) = \log \left(-
   \frac {\za \,\foa(T/2; \theta,  z)}{\left[ f_1(\theta, z) k_1(\theta, 0)
     \right ]^{1/2}}\right),\\
   f_1(\theta, z) =  - 
 \frac {a^3} 2
  \sum_{\vec x} \avg{\zetabar_1\,\gamma_5\,\zeta_3
    \zetabar_3'\,\gamma_5 \,\zeta_1'},\quad
  k_1(\theta, z) =  - 
 \frac {a^3} 2
 \sum_{\vec x} \avg{\zetabar_1\,\gamma_1\,\zeta_3
    \zetabar_3'\,\gamma_1 \,\zeta_1'}.
 \end{gather}
The renormalization constants $\zv,\za$ are fixed by Ward identities
\cite{Luscher:1996jn,Maiani:1986yj}. The quark masses are renormalized
using the lattice minimal subtraction scheme
\cite{Sint:1997jx,Gabrielli:1990us} at one-loop order,
\begin{gather}
  \label{eq:1}
  \mr = \zmlat(g_0^2, a\mu) \mq \,\of{ 1 + a\, b_m(g_0^2)\, \mq }, 
  \quad \mq = m_0 - \mc,\\
  \bm = - 0.5 - 0.07217(2)\,C_F\, g_0^2,\quad
  \zmlat(g_0^2, a\mu) = 1 - \frac  1 {2\,\pi^2} \log (a\mu)  g_0^2.
\end{gather}
\subsection{Static approximation}
We can use the symmetries of the static theory to establish a relation
between the renormalized, renormalization group invariant (RGI) axial
vector and vector currents \cite{Sommer:2010ic},
\begin{align}
  \label{eq:186}
  V_0^\hqet = \CP(M_b/\Lambda_\msbar)\, Z^\stat_{\mathrm{A},\rgi} (g_0)\,
  \zva^\stat(g_0) \,V_0^\stat, \\
  \label{eq:187}
  A_k^\hqet = \CV(M_b/\Lambda_\msbar)\, Z^\stat_{\mathrm{A},\rgi} (g_0)\,
  \zva^\stat(g_0) \,A_k^\stat .
\end{align}
At one loop level, we have in the lattice minimal subtraction scheme
\begin{gather}
  \label{eq:165}
  (V_\mathrm{lat}^\stat)_0(\mu) =
  Z^\stat_\mathrm{A,lat}(\mu)\,\zva^\stat V^\stat_0,\quad
  (A_\mathrm{lat}^\stat)_k(\mu) =
  Z^\stat_\mathrm{A,lat}(\mu)\,\zva^\stat A^\stat_k,\\
  Z^\stat_\mathrm{A,lat}(\mu) = 1 - \gamma_0 \log(a\mu)\, g_0^2 +
  O(g_0^4), \quad
  \zva^\stat = 1 + \left( \zva ^\stat \right) \ord 1
  g_0^2 + O(g_0^4),
\end{gather}
with (c.f.\ \cite{Palombi:2007dt,Shifman:1986sm,Politzer:1988wp})
\begin{gather}
  \left( \zva ^\stat \right) \ord 1 = 0.0521(1),\quad  \gamma_0 = - \frac 1 {4\pi^2}.
\end{gather}
We define the static counterparts to \phia\ and \phiv, 
\begin{equation}
  \label{eq:2}
    \xa(\mu) =
    \log\left(Z^\stat_\mathrm{A,lat}(\mu)\,\zva^\stat 
      \right) + \xabare,\quad 
    \xv(\mu) = \log \left(
      Z^\stat_\mathrm{A,lat}(\mu)\,\zva^\stat\right) + \xvbare, 
\end{equation}
with \xabare\ and \xvbare given (without the renormalization factors)
as in (\ref{eq:247}) with the heavy quark flavor replaced by a static
one.  We may then expect that at one loop level
\begin{equation}
  \label{eq:3}
    \phiv(L,\mr) = \Ba g_0^2 + X_\mathrm{V}(\mr) + O(1/\mr),\quad
  \phia(L,\mr) = \Bv g_0^2 + X_\mathrm{A}(\mr) + O(1/\mr),
\end{equation}
where we have from \cite{Sommer:2010ic}, using the ratio of $\CV$ and
$\CP$,
\begin{equation}
  \label{eq:4}
  \Ba = -0.137(1),\quad \Ba - \Bv = 0.016900.
\end{equation}


\section{Results}
\subsection{Tree level}
\label{sec:tree-level-1}

\begin{figure}[ht]
  \centering
  \includegraphics{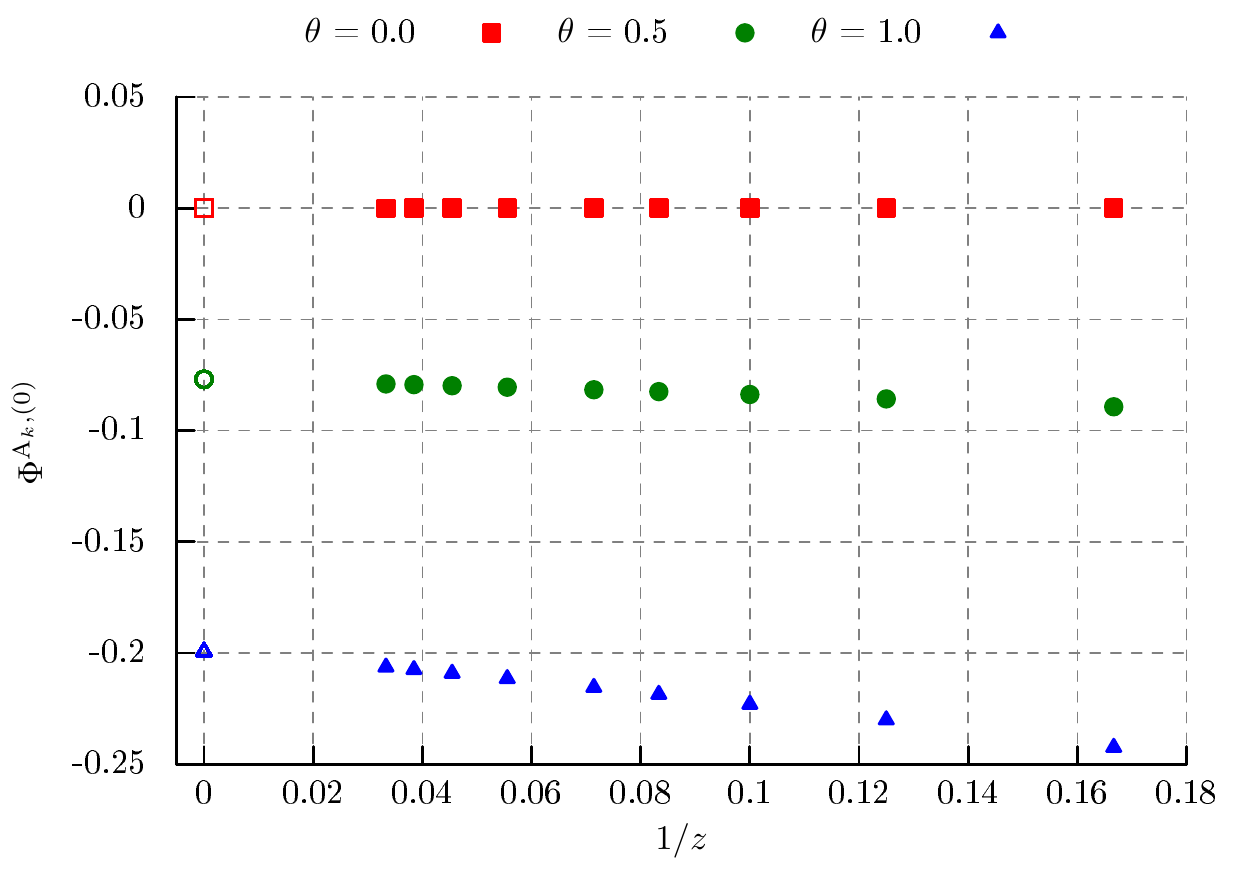}
  \includegraphics{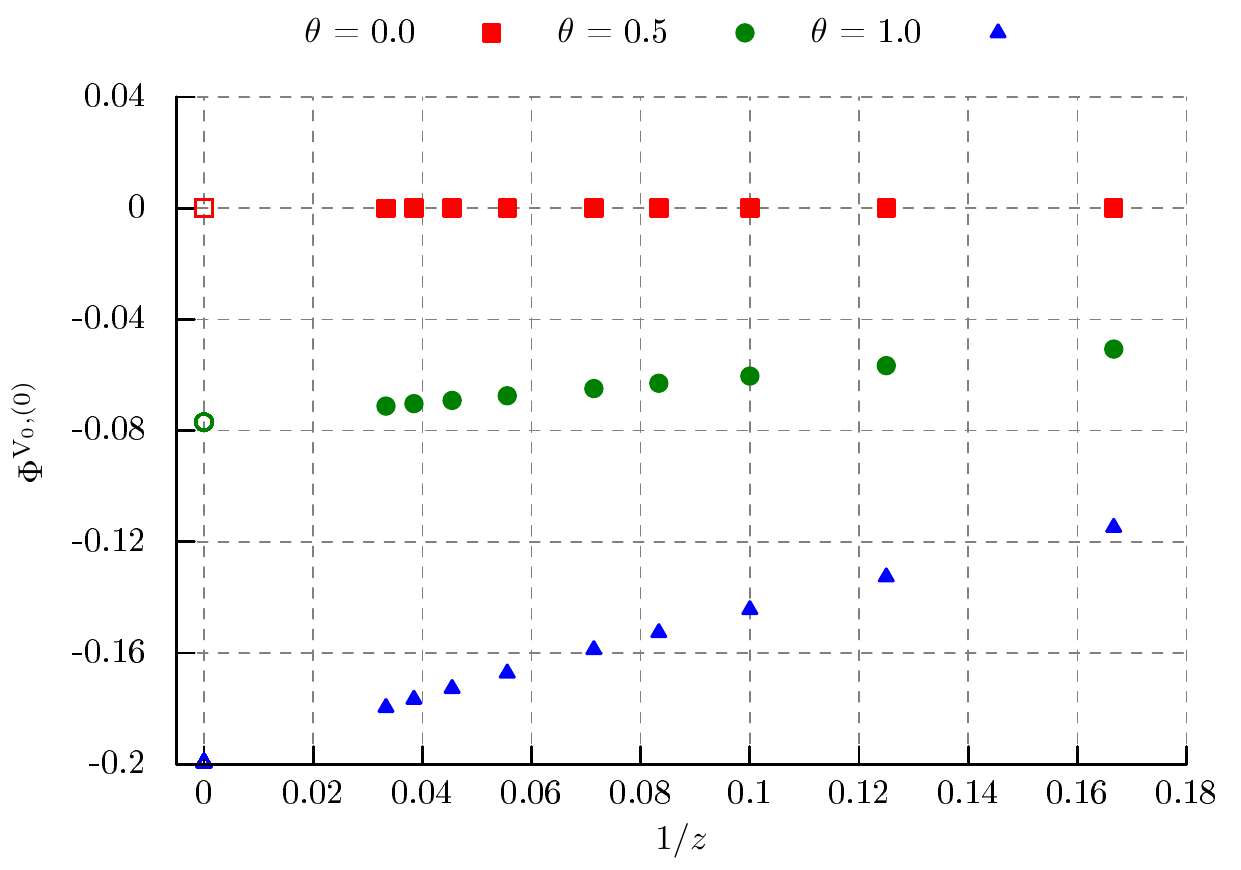}
  \caption
  {\phia\ and \phiv\ at tree level in the continuum limit.  The point size
  is bigger than the error.}
  \label{fig:tree}
\end{figure}

Tree level results for \phiv\ and \phia\ are shown in figure
\ref{fig:tree}, the points at $1/z = 0$ correspond to the static values
$\xao 0$ and $\xvo 0$. For $\theta = 0.5,
1.0$, the dependence on $1/z$ is linear with a small slope. In the
interesting region of $z = 10$, which is the typical matching point
for B physics used in \cite{Morte:2006cb}, the $1/z$ corrections at
tree level are of a few percent. For $\theta = 0$, there is no
dependence on $z$ and $L/a$ at tree level for both observables.  The
continuum limit at tree level was extracted from computations with
$L/a$ up to 200, using the fitting procedure explained in
\cite{Bode:1999sm}.

\subsection{One loop}
\label{sec:one-loop-1}

\begin{figure}[ht]
  \centering
  \includegraphics{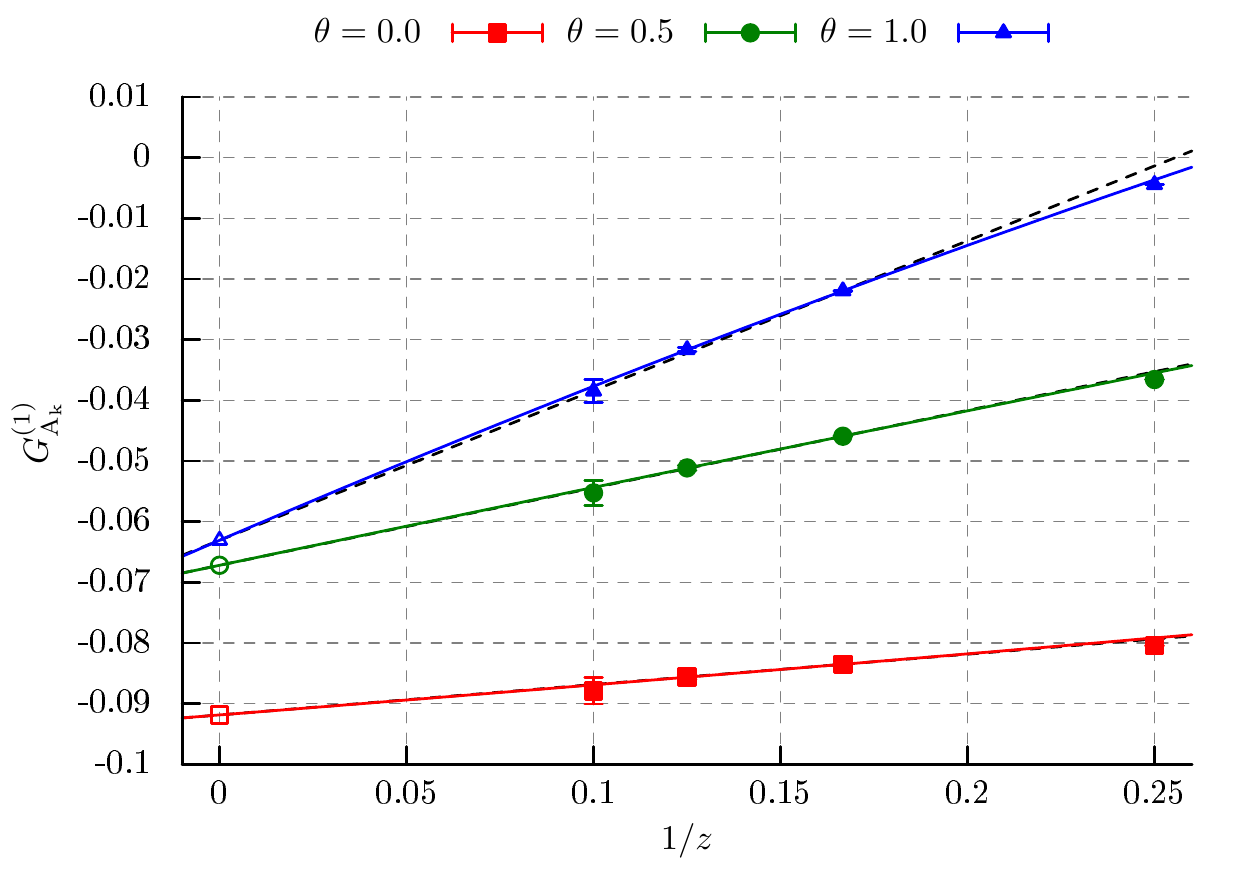}
  \includegraphics{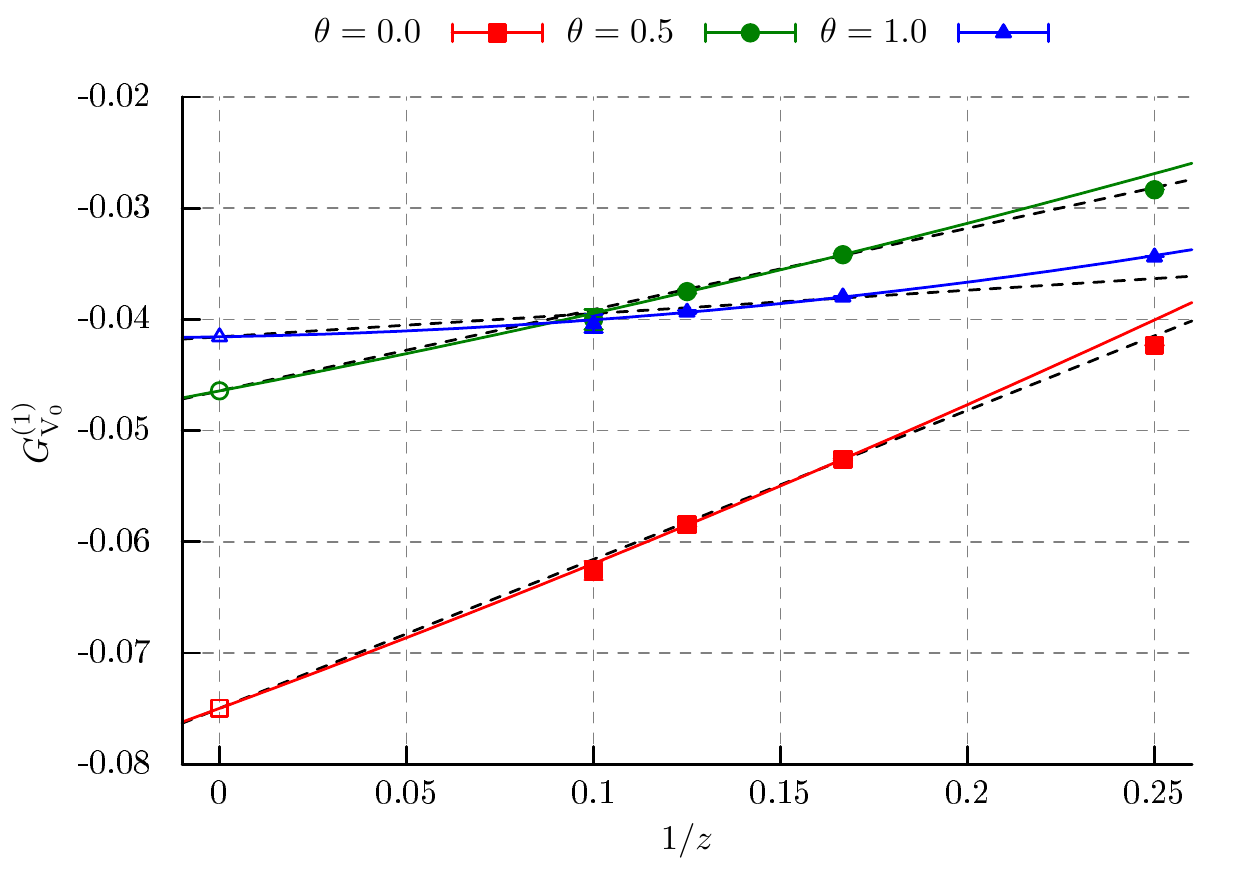}
  \caption{\gao 1 and \gvo 1 in the continuum limit. The
    dashed lines are a linear fit in $1/z$, setting $c_2 = 0$.  The
    data points at $z = 4$ are not included in the fits.}
  \label{fig:loop}
\end{figure}

The observables $f_1$, \phiv, and \phia\ were calculated at one loop
level for $z = 4,6,8,10$ (and at $z = 0$ for $f_1$, $k_1$), with
lattice resolutions up to $L/a=40$, and for $\theta \in \{ 0.0, 0.5,
1.0\}$. Furthermore, we evaluated the static counterparts $f_1^\stat$,
\phivs\ and \phias\ for lattices with $L/a$ up to 28. No bigger
lattice sizes are required for the HQET quantities, since their
continuum limit is easier to obtain due to a weaker
$a/L$-dependence. The continuum values are presented in figure
\ref{fig:loop}, the points for $1/z = 0$ are again the static values
on which we will comment on shortly. The extraction of the continuum
values is done with the method mentioned above. To establish the
connection to the static limit, we define the one loop quantities
\begin{align}
  \galph(z) = &\;\Phi^{\alpha,(1)} (z) + \gamma_0 \log(z)
  \xrightarrow{1/z \to 0} \halph, \quad \alpha \in \{V_0, A_1\},\\
  \hvo 1 =  &\;\xvbareo 1  +
    \left(\zva^\stat \right)\ord 1  + \Ba - \gamma_0 \log(a/L),\\
  \hao 1 =  &\;\xabareo 1  + \left(\zva^\stat \right)\ord 1  +
  \Bv - \gamma_0 \log(a/L).
\end{align}
The fit functions included in the plots are of the form $\galph(z) =
\halph + c^\alpha_1/z + c^\alpha_2/z^2 $. Even though they are only
indicative (since possible terms of the form $\log(z) / z^n$ are not
included), one can anticipate that the higher order corrections
corresponding to $c_2$ are rather small. We did not attempt to fit the
logarithmic terms due to the small number of available data points.

In the case of \phia, a value of $\theta = 0$ seems to minimize the
higher order corrections in $1/z$ both at tree level and at one
loop. For \phiv, the value $\theta = 0.5$ seems to be a good
compromise considering the corrections at the two orders in
perturbation theory we investigated. One should always keep in mind
that the small observed dependence on $1/z$ will be eliminated once
the $1/\mh$ corrections are included in the effective theory. Finally
only the $1/z^2$-terms, manifesting themselves in the curvature of the
fits, will remain as corrections.

%


\section{Conclusions}

We have seen that the proposed observables for the matching are
dominated by the lowest orders in $1/m$ and hence one would strongly
suggest to use them in a non-perturbative study. For the phase angle
$\theta$, we would suggest the value $0.5$ as a compromise to minimize
$1/z$-effects at one-loop and tree level, but the current study does
not indicate a strong preference for this value. A further
investigation including the matching conditions for all 19 parameters
is currently carried out within the ALPHA collaboration.

We could also demonstrate the usefulness of the \pastor\ software
package for applications beyond the classical realms of lattice
perturbation theory such as renormalization and improvement. The
publication of an in-depth description of \pastor\ along with its
source code is planned for the near future.

\section{Acknowledgments}

The calculations were carried out on the PC farm at DESY Zeuthen. This
work has been supported by the DFG Sonderforschungsbereich TR9
Computergest\"utzte Theoretische Teilchenphysik and the Research
Executive Agency (REA) of the European Union under Grant Agreement
number PITN-GA-2009-238353 (ITN STRONGnet).


\appendix
\section{Plots}


\end{document}